\documentclass[prd,twocolumn,nofootinbib,preprintnumbers
]{revtex4-1}

\usepackage{natbib}
\usepackage{amssymb,amsbsy,amsmath,amsfonts}
\usepackage{mathrsfs}
\usepackage{graphicx}
\usepackage{epsf,epsfig,float,latexsym,amsthm,fancyhdr,rotating}
\usepackage{graphics,psfrag,longtable}
\usepackage{slashed}
\usepackage{esint}
\usepackage{nicefrac}
\usepackage{braket}
\usepackage{mathtools}

\newcommand{\nn}{\nonumber}

\renewcommand{\eth}{\partial}
\newcommand{\be}{\begin{equation}}
\newcommand{\ee}{\end{equation}}

\usepackage{cmbright}		

\usepackage[colorlinks,citecolor=blue,linktoc=all,linkcolor=cyan]{hyperref}

\begin{document}

\title {Comments and extensions of a  suggestion for a relativistic charge density definition}

\author{Carl E. Carlson}

\affiliation{Physics Department, William \& Mary, Williamsburg, Virginia 23187, USA}

\begin{abstract}
A recent suggested definition of a relativistically correct three dimensional charge density of an extended hadron is shown to be physically and intuitively connected to an earlier relativistically correct two dimensional charge density studied in the context of light-front physics.  Looking at spin-1/2 hadrons, such a connection is shown to exist for both the polarized and unpolarized cases.
\end{abstract}
\date{August 21, 2023
}
\maketitle


\section{Introduction}



The well-known nonrelativistic connection that gives a hadron's charge density as the Fourier transform of the charge form factor is not justified relativistically.  The question becomes how one could obtain a relativistically correct charge density from the lepton scattering data.  A by now older suggestion~\cite{Miller:2007uy,Burkardt:2002hr} is to work in the light front formalism, or in an infinite momentum frame, with the hadron and virtual photon traveling in essentially opposite directions and where one can think of the hadron as flattened in its direction of motion.  It is then possible to obtain a relativistically well defined area charge density by projection onto the transverse plane, both for polarized and unpolarized situations.  

One may still seek a relativistically correct way to determine a three dimensional charge density from form factors.  Ref.~\cite{Epelbaum:2022fjc} makes an interesting suggestion, with followups in~\cite{Panteleeva:2022khw,Panteleeva:2022uii,Panteleeva:2023evj}.  The goal of this note is to, after a brief review, understand the charge density definition of~\cite{Epelbaum:2022fjc,Panteleeva:2022khw} in terms of its connection to the older two dimensional relativistic densities and to do so for both unpolarized and polarized hadrons. Other recent discussions relevant to charge density expressions are in~\cite{Freese:2021mzg,Jaffe:2020ebz,Lorce:2022cle,Guo:2021aik}. 

To anticipate the physical situation, both the three dimensional and two dimensional charge density definitions involve, if expressed in momentum space, hadrons moving extremely fast and Lorentz flattened to a disk.  The 3D density is then the density on a disc averaged over all orientations.  The charge on the disc in a ring of circumference $2\pi r$ and a certain thickness averages, as the oreintation varies, to the charge on a sphere of the corresponding thickness and surface $4 \pi r^2$.  Hence the relation between the newly defined 3D density $\rho_1(r)$ and the surface charge density on the disc, the older light front projected density $\sigma_1(r)$, should be
\be         \label{eq:anticipate}
\rho_1(r) = \frac{ \sigma_1(r) }{ 2r } .
\ee
We shall see how this anticipation, and its polarized counterpart, fares as we proceed.


\section{Review}


To define a charge distribution one begins with a statement of what hadron state is to be used.  The suggestion in~\cite{Epelbaum:2022fjc}, here with hadron helicity $\lambda$ included, is equivalent to a state written as~\cite{Miller:2007uy,Burkardt:2002hr,Epelbaum:2022fjc,Panteleeva:2022khw}
\be             \label{eq:statedef}
    \ket{\Phi, \lambda} = \mathcal N
\int \frac{ d^3p }{ (2\pi)^3 } 
\frac{1}{ \sqrt{2 E_p} } 
\ket{\vec p,\lambda }   \,.
\ee
The states within the integral are momentum eigenstates for the extended hadron where $\vec p$ is the total momentum and the normalization is
\be
\braket{ \vec p{\,'}, \lambda' | \vec p, \lambda } = (2\pi)^3 2E_p \delta_{\lambda'\lambda}
\delta^3( \vec p{\, '} - \vec p \, )  ;
\ee
$E_p = \sqrt{ m^2 + \vec p^2}$, where $m$ is the hadron's mass, and
\be                     \label{eq:norm}
    | \mathcal N |^2
\int \frac{ d^3p }{ (2\pi)^3 } = 1  .
\ee
One could include a wave function or profile function in the integral, with a limiting procedure to ensure the momenta are dominantly very large magnitude.  This is done in~\cite{Epelbaum:2022fjc}, but the manipulations and the physical genesis of the results are clear without the extra notation.  If done, however, the results can be shown to be the same as will be found here. 

The momentum distribution that makes up the state $\ket{\Phi,\lambda}$ is de facto spherically symmetric, which is true also in~\cite{Epelbaum:2022fjc}.  The state is arguably a state with no overall momentum, centered at the origin.  A discussion of how to construct center of mass coordinate operators and eigenstates is placed in the Appendix, thought this is not strictly crucial for what follows.  What is important is that, when we want to specify the charge density and radius of a state,  we are choosing to do it for a state that is stationary and centered at the origin, $\vec R = 0$.  A state with such a definite location requires very high momentum components in its momentum space description, and the dominance of high momentum components in the state is crucial in what follows.



The charge density is defined from
\be             \label{eq:density}
\rho_{\lambda'\lambda}
    = \braket{ \Phi, \lambda' | \,
        j^0(\vec r) | \Phi, \lambda },
\ee
where $j^0$ is the density component of the electromagnetic current and time $t=0$ is tacit.  We will work out the charge densities for the spin-1/2 states
\be                 \label{eq:spinstate}
\ket{ \Phi, s} = \frac{1}{ \sqrt{2} }
    \left( \ket{ \Phi, \lambda = 1/2}
    + e^{i\phi_s}\ket{ \Phi, \lambda = -1/2} \right),
\ee
somewhat differently from the tack taken in~\cite{Panteleeva:2022khw}.

As a further reminder, the analogous light front treatment that leads to the 2D areal charge distribution begins with hadron states
\be
    \ket{ p^+, \vec R_\perp = 0, \lambda}
    = \mathcal N' \int
    \frac{d^2p_\perp}{(2\pi)^2}
    \ket{p^+,\vec p_\perp, \lambda },
\ee
where $\lambda$ is now the light front helicity and $p^+ = p^0 + p^z$.  The projected surface charge densities come from
\begin{align}
    &\sigma_{\lambda'\lambda}(\vec r_\perp)
    =  1/(2p^+) \ \times \nn\\
&   \hskip 0.7 em
\braket{ p^+, \vec R_\perp = 0, \lambda' |\,
    j^+(0,r^-,\vec r_\perp) |
    p^+, \vec R_\perp = 0, \lambda }   ,
\end{align}
where $r^+ = t+z =0$. For spin-1/2 hadrons, the results are worked out for~\cite{Carlson:2007xd,Miller:2010nz} 
\begin{align}
&\ket{p^+,\vec p_\perp, s } = 
    \frac{1}{ \sqrt{2} }  \ \times  \nn\\
&   \hskip 1.7 em
\left( \ket{p^+,\vec p_\perp, \lambda=1/2 }
    + e^{i\phi_s}
    \ket{p^+,\vec p_\perp, \lambda=-1/2 } \right),
\end{align}
One finds~\cite{Carlson:2007xd,Miller:2010nz}
\be
\sigma(\vec r_\perp) =
    \sigma_{ss}(\vec r_\perp)=
    \sigma_1(r_\perp)
    + \sin(\phi_r - \phi_s)
    \sigma_2(r_\perp)   ,
\ee
where $r_\perp = | \vec r_\perp |$, $\phi_r$ is the azimuthal angle of $\vec r$, and
\begin{align}           \label{eq:lfresults}
    \sigma_1(r) &= \int \frac{ Q dQ }{2\pi}
        J_0(Qr) F_1( -Q^2), \nn\\
    \sigma_2(r) &= \int \frac{ dQ }{2\pi}
        \frac{ Q^2 }{ 2m }
        J_1(Qr) F_2( -Q^2).
\end{align}
The $J_\nu$ are Bessel functions and $F_1$ and $F_2$ are the Dirac and Pauli form factors, respectively.

We should comment that the light front calculation works for any allowed value of $p^+$.  The hadron's longitudinal momentum could be very fast, in which case the hadron will appear Lorentz flattened and it is natural to speak of a 2D charge distribution on a disc. Or the hadron could be at rest and hence appear quite three dimensional.  In this case one still gets a 2D charge distribution, with the projection bei  ng done by the photon beam.  The 2D charge distribution is the same for all $p^+$, and it is convenient to think of it as a charge density on a disc.


\section{3D development}


The 3D density, Eq.~\eqref{eq:density}, is
\begin{align}
    \rho_{\lambda' \lambda}(\vec r) &=
    | \mathcal N |^2 \int \frac{d^3P d^3q}
    { (2\pi)^6 \sqrt{2E_p' 2E_p} } 
    e^{-i \vec q \cdot \vec r} \times \nn\\
&\quad 
\braket{ \vec P + \frac{\vec q}{2},\lambda'
    |  \, j^0(0)  | 
    \vec P - \frac{\vec q}{2},\lambda }.
\end{align}
Note that $P = | \vec P |$ is almost always large, de facto $P \to \infty$, while $|\vec q\,|$ is limited by form factors.  We can omit many corrections of 
$\mathcal O(|\vec q|/P)$, so that for example 
$\sqrt{2E_p' 2E_p} \to 2E_p$.

The matrix element is
\begin{align}
&\braket{ \vec P + \frac{\vec q}{2},\lambda'
    |  \, j^0(0)    \,  | 
    \vec P - \frac{\vec q}{2},\lambda } =
\bar u(\vec P + \frac{\vec q}{2},\lambda')
    \ \times
            \nn\\
&\hskip 2 em
    \left[ \gamma^0 G_M(q^2) - 
    \frac{P^0}{m} F_2(q^2) \right] 
    u(\vec P - \frac{\vec q}{2},\lambda),
\end{align}
where $G_M = F_1 + F_2$, and the Dirac spinors are
\begin{align}
    u(\vec p,\lambda) = \frac{1}{\sqrt{E_p+m}}
    \left( E_p +m + \vec\alpha\cdot\vec p \,
    \right) \chi_\lambda(\hat p).
\end{align}
The $\chi_\lambda$ are helicity eigenstates that have only upper components, and the $\vec\alpha$ are standard Dirac matrices. After some manipulation
\begin{align}
    \rho_{\lambda' \lambda}(\vec r\,) &=
    \int \frac{d^2 \hat P }{ 4 \pi }
        \frac{ d^3q }{ (2\pi)^3 }
    e^{-i \vec q \cdot \vec r} \times \nn\\
&\quad 
\left[ \delta_{\lambda'\lambda} F_1(q^2) +
\frac{i}{2m} \vec\sigma_{\lambda'\lambda}
    \cdot \vec q \times \hat P \  F_2(q^2)
        \right] .
\end{align}
Here,
\be
    \vec\sigma_{\lambda'\lambda} =
    \chi_{\lambda'}^\dagger(\hat P) \,  
    \vec \sigma \,
    \chi_\lambda(\hat P)    \,,
\ee\\
we have let $d^3P = P^2 dP \, d^2 \hat P$ and done the $dP$ integral using the normalization condition, Eq.~\eqref{eq:norm}.  Terms of 
$\mathcal O(|\vec q|/P)$ and 
$\mathcal O(m/P)$ are omitted.  The arguments of the form factors become, as in~\cite{Epelbaum:2022fjc},
\be
    q^2 = (E_p' - E_p)^2 - \vec q^{\,2} =
( \hat P \cdot \vec q \, )^2 - \vec q^{\,2}=
    - q_\perp^2 \,,
\ee
where $\vec q_\perp$ is the part of $\vec q$ that is perpendicular to $\hat P$.

To evaluate using the spin states 
$\ket{ \Phi, s}$, defined in Eq.~\eqref{eq:spinstate}, let
\begin{align}
    \chi_s &= \chi_s(\hat P) = 
    \frac{1}{ \sqrt{2} } \left( 
    \chi_{1/2}(\hat P) + \chi_{-1/2}(\hat P)
    \right) \,,
\end{align}
and
\begin{align}
    \vec \sigma_{ss} &= 
    \vec \sigma_{ss}(\hat P) =
    \chi_s^\dagger \ 
    \vec\sigma \ \chi_s  \equiv \hat s.
\end{align}
To evaluate $\hat s$, recall that the $\chi_\lambda(\hat P)$ are defined using the Jacob-Wick phase convention.  In this convention, states defined with spin projection $\lambda$ along the $z$-axis are rotated to the required direction as
\be
    \chi_\lambda(\hat P) = R_z(\phi_P) R_y(\theta_P) R_z^{-1}(\phi_P) \,  \chi_\lambda(\hat z) 
    \equiv R_P \, \chi(\hat z)
\ee
where the direction $\hat P$ is defined by the polar and azimuthal angles $\theta_P$ and  $\phi_p$.  The direction of $\hat P$ itself can be given as
\be
\hat P = R_P \hat z,
\ee
with the protocol that $R_P$ is represented by $2\times 2$ or $3 \times 3$ matrices, as the situation demands.  In like fashion, $\hat s$ is a vector and,
\be
\hat s = R_P \hat s_0   \,,
\ee
where $\hat s_0$ is a base value evaluated with states quantized in the $z$ direction,
\be
    \hat s_0 = \chi_s(\hat z) \, \vec\sigma
    \chi_s(\hat z) =
    \hat x \cos\phi_s + \hat y \sin\phi_s
    \,.
\ee

The density is now
\begin{align}
    \rho(\vec r) &= \rho_{ss}(\vec r)
    = \int \frac{d^2 \hat P }{ 4 \pi }
        \frac{ d^3q }{ (2\pi)^3 }
    e^{-i \vec q \cdot \vec r} \times \nn\\
&\quad 
\left[ F_1(-q_\perp^2) +
\frac{i}{2m} \hat s \cdot \vec q 
    \times \hat P \  F_2(-q_\perp^2)
        \right] .
\end{align}
The $d^3q$ integrals can be done while holding $\hat P$ fixed.  Split both $\vec q$ and $\vec r$ into components parallel and perpendicular to $\hat P$.  For the first integral,
\begin{align}
    &\int \frac{ d^3q }{ (2\pi)^3 }
    e^{-i \vec q_\perp \cdot \vec r_\perp - i q_\parallel r_\parallel} 
    F_1(-q_\perp^2)     \nn\\
&=  \delta(r_\parallel) 
    \int\frac{ d^2q_\perp }{ (2\pi)^2 }
    e^{-i \vec q_\perp \cdot \vec r_\perp}
    F_1(-q_\perp^2)     \nn\\
&=  \delta(\hat P \cdot \vec r)  \ 
        \sigma_1(r_\perp)   \,,
\end{align}
where $\sigma_1(r_\perp)$ is indeed the same function seen when reviewing the projected surface charge densities in the light front case, Eq.~\eqref{eq:lfresults}.  For the second integral, noting that $\vec q$ can be replaced by $\vec q_\perp$,
\begin{align}
    &\frac{i}{2m} 
    \int \frac{ d^3q }{ (2\pi)^3 }
    \vec q_\perp 
    e^{-i \vec q_\perp \cdot \vec r_\perp}
    F_2(-q_\perp^2) \nn\\
&=  \frac{1}{2m} \,\delta(\hat P\cdot\vec r)
    \, \vec\nabla_\perp 
    \int\frac{ d^2q_\perp }{ (2\pi)^2 }
    e^{-i \vec q_\perp \cdot \vec r_\perp}
    F_2(-q_\perp^2) \nn\\
&=  \frac{1}{2m} \,\delta(\hat P\cdot\vec r)
    \, \vec\nabla_\perp
    \int \frac{ Q dQ}{2\pi}  J_0(Q r_\perp)
    F_2(-Q^2)   \nn\\
&=  - \delta(\hat P\cdot\vec r) \, \hat r \,
    \sigma_2(r) \,,
\end{align}
where $Q = | \vec q_\perp$ and we replaced $r_\perp$ by $r$, which is allowed since the delta-function sets $r_\parallel$ to zero.  Function $\sigma_2(r_\perp)$ is the same as in the spin dependent term in the light front case, Eq.~\eqref{eq:lfresults}.

Now,
\begin{align}
    &\rho(\vec r)
    = \int \frac{d^2 \hat P }{ 4 \pi r }
    \delta(\hat P\cdot \hat r)
\left[ \sigma_1(r) + 
\hat s \cdot \hat r 
    \times \hat P \  \sigma_2(r)
        \right] .
\end{align}

The integral for the first term is elementary, and gives the result anticipated by picturing the physical situation, recorded in Eq.~\eqref{eq:anticipate}.
In the second term, we remember that $\hat s$ depends on $\hat P$.  Also in the second term, using $d^2\hat P = d\phi_p d(\cos\theta_p)$ and integrating the $\delta$-function using $d(\cos\theta_p)$, leads to a Jacobian
\be
J = \left| \frac{\eth(\hat P \cdot \hat r)}
    {\eth(cos\theta_p)} \right|
=   \left| \cos\theta_r \right|  \,
( 1 + \tan^2\theta_r cos^2 \phi_p) .
\ee
The azimuthal integration can then be done straightforwardly, yielding
the final result
\begin{align}
    \rho(\vec r)
    = \rho_1(r) + 
    \rho_2(r) \sin(\phi_r-\phi_s)   \,
\frac{  \sin \theta_r  }{ 1 + | \cos\theta_r | }   ,
\end{align}
with
\be
    \rho_{1,2}(r) = \frac{ \sigma_{1,2}(r) }
    {2r}    \,.
\ee


\section{Closing remarks}


The relativistic definition of the 3D charge density given in~\cite{Epelbaum:2022fjc}, and the ensuing manipulations to express that charge density in terms of the form factor measured in lepton hadron scattering, has good features.  It is the expectation value of the charge density operator evaluated for an extensive hadron in a particular state, which one may may be able to separately argue is a position eigenstate.  

The state for which the charge density has been evaluated is, in momentum space, the superposition of states of unlimited momentum magnitude traveling in all directions.  That nearly all the momenta are much much larger than the mass means that the mass of the state plays no kinematic role in the relation between the charge density and the  form factors, and appears only as a normalizing factor in the definition of $F_2$.

As nearly all the momenta are very large, for a given direction of momentum the hadron appears like a Lorentz flattened pancake with its cross section perpendicular to the momentum direction. The charge density on the surface of that pancake is the same as earlier considerations in a light front formalism, which can be thought of as hadrons moving fast in a given direction.  In the light front case, the areal charge density in realistically well defined.  Intuitively, the 3D charge density now under discussion would be the same as disc with this areal, or 2D, charge density rotated in all possible directions to fill a 3D volume, and averaged over.  This view leads to the relation between the 3D and 2d charge densities given earlier, Eq.~\eqref{eq:anticipate}, and verified by detailed calculations.

For the polarized hadron, analogous results follow.  Polarization means there is a definite directional vector associated with a stationary hadron,  allowing for direction position dependence of the charge density.  The basic function giving the position dependence in the 3D polarized case has the same relation to its 2D counterpart as in the unpolarized case.

Hence there is a definite connection of the newly suggested 3D relativistic charge density to the 2D relativistic charge density known earlier, and the intuitive nature of the connection could be viewed as positive support for the new 3D suggestion.


\section*{Acknowledgements}
I thank Jerry Miller for useful (email) conversations and thank the National Science Foundation (USA) for support under grant PHY-1812326.



\appendix
\section{CM Coordinate Operator}


If one has eigenstates of total momentum, one has also a total momentum operator, one can define coordinate eigenstates and a coordinate operator whose components have the correct canonical commutation relations. We shall demonstrate this below.  The coordinate being conjugate to the total momentum is what is standardly called the center of mass coordinate.

The situation considered here, just finding a center of mass coordinate where there are momentum eigenstates, is different from trying to define relativistically a CM coordinate from constituent coordinates in a multicomponent system.  Pryce, in a well known paper~\cite{Pryce:1948pf}, pointed out at least two difficulties that can occur in this case.  One is that in a multicoordinate system, the results of the definitions may be frame dependent.  Another is that the coordinate components don't commute with each other.  Pryce showed sample definitions that avoided one or the other of these dificulties, but no definition that avoided both.

To return to the present task, study for simplicity the single particle sector in the scalar case, with the extended objects having momentum eigenstates $\ket{ \vec p \,}$ normalized by
\be
\braket{ {\vec p\,}' | \vec p \, } = (2\pi)^3 2E_p \delta^3( \vec p\,' - \vec p \,).
\ee
The identity operator and momentum operator can be given by
\begin{align}
    I &=  \int \frac{d^3p}{ (2\pi)^3 2E_p } \ket{\vec p \,} \bra{ \vec p \,}    \,,      \nn\\
\vec P &=  \int \frac{d^3p}{ (2\pi)^3 2E_p } \ket{\vec p\,}  \vec p \, \bra{\vec p \,}      \,,
\end{align}
and it is easy to verify 
$I \ket{ \vec p \,} = \ket{ \vec p \, }$ and 
$\vec P \ket{ \vec p \,} = \vec p \,
\ket{ \vec p \, }$.

Define coordinate states 
\be
\ket{ \vec x \,} =  \int \frac{d^3p}
{ (2\pi)^3  \sqrt{ 2E_p}} 
e^{-i \vec p \cdot \vec x}   \ket{\vec p \,}	.
\ee
These have properties
\begin{align}
\braket{\vec x\,' | \vec x \,} 
&= \delta^3 (\vec x\,' - \vec x \,)	,	\nn\\
\braket{\vec p \, | \vec x \, } 
&= \sqrt{ 2E_p } e^{-i \vec p \cdot \vec x}		.
\end{align}
The identity operator can also be given by
\be
I = \int d^3x \, \ket{\vec x \,} \bra{\vec x \,},
\ee
and the position operator is defined by
\be
\vec X = \int d^3x  \, 
\ket{\vec x \,}  \vec x \, \bra{\vec x \,}	.
\ee
It is easy to test 
$I \ket{\vec x\,} = \ket {\vec x\,}$ and 
$\vec X \ket{\vec x \,} = \vec x \,
\ket {\vec x \,}$.

The position operator can also be written
\be
\vec X = \int d^3x
\frac{d^3p}{ (2\pi)^3 \sqrt{2E_p} }  
\frac{d^3p'}{ (2\pi)^3 \sqrt{2E'_p} } 
e^{i( \vec p - \vec p' \,) \cdot \vec x }
	\ket{\vec p\,'} \vec x \, \bra{\vec p \,}	. 
\ee

It remains to consider the commutator,
\begin{align}
&X_i P_j - P_j X_i =    \nn\\
&=\int d^3x \, \frac{d^3p}{ (2\pi)^3 }  \frac{d^3p'}{ (2\pi)^3 } \,
	\ket{\vec p'\,} x_i ( p_j - p'_j ) 
    \bra{\vec p \,}	\, 
    e^{i(\vec p - \vec p') \cdot \vec x }	\nn\\
&=	 \int d^3x \, \frac{d^3p}{ (2\pi)^3 }  \frac{d^3p'}{ (2\pi)^3 } \,
	\ket{\vec p\,'}  \bra{\vec p \,}  
    x_i (-i \frac{ \eth }{ \eth x_j } )	\, 
    e^{i(\vec p - \vec p') \cdot \vec x }   .
\end{align}

It is crucial that a space derivative does not affect the momentum state,
\be
\frac{ \eth }{ \eth x_j } \ket{\vec p \,} = 0 .
\ee
This is on one hand natural.  However, should it not be true at first instance, we prove as a separate lemma later that the phase can always be chosen to make it true.

Integration by parts then gives
\begin{align}
X_i P_j - P_j X_i 
=	i \delta_{ij}		\int  \frac{d^3p}{ (2\pi)^3 }		 \ket{p}  \bra{p}		,
\end{align}
or 
\be
\left[  X_i , P_j  \right] = i \delta_{ij}  I  ,
\ee
the canonical result.

Regarding the promised lemma,  it seems natural that the momentum eigenstate should have no parametric dependence on position, and 
$(\eth/\eth x_i) \ket{\vec p\,} = 0$ follows accordingly. On the other hand, the state $\ket{\vec p\,}$ multiplied by a function of position is still a momentum eigenstate, and then this condition won't hold, so we pose the question whether we can generally find momentum eigenstates with 
$(\eth/\eth x_i) \ket{\vec p\,} = 0$ if we start with states 
$\ket{\vec p\,}_a$ where $(\eth/\eth x_i) \ket{\vec p\,}_a \ne 0\,$?   We will allow use of normalization and uniqueness (that is, there is only one independent momentum eigenstate for each momentum).
So we have

\paragraph*{Lemma:}  The phase of the momentum eigenstates can always be chosen so that 
\be
\frac{ \eth }{ \eth x_j } \ket{\vec p \,} = 0 .
\ee

\paragraph*{Proof:} The space derivative commutes with the momentum operator, so the derivative must yield a momentum eigenstate with the same momentum, and by uniqueness the same eigenstate, up to a possible position dependent factor,
\be
\frac{\eth}{\eth x_i} \ket{\vec p\,}_a      = i \alpha_i(\vec x\,)  
        \ket{\vec p\,}_a	.
\ee
We can further prove that 
$\alpha_i(\vec x\,)$ is real (from the normalization condition).

Since there are three directions,
\be
\vec\nabla \ket{\vec p\,}_a = 
\big(   i \alpha_1(\vec x\,), 
        i \alpha_2(\vec x\,), 
        i \alpha_3(\vec x\,) \big) 
        \ket{\vec p\,}_a 
	\stackrel{\text{def}}{=}   i  
\vec u\,(\vec x\,)  \ket{\vec p\,}_a.
\ee 
Note that $\vec\nabla \times \vec u = 0$, which follows from letting 
$\vec\nabla\times\vec\nabla$ act on 
$\ket{\vec p\,}_a$.

Seek a state 
$\ket{p} = e^{i\phi(\vec x)}  \ket{p}_a$ satisfying the zero derivative requirement.  The transformation is just a phase, for normalization maintenance.  Then
\begin{align}
0 &= \vec\nabla \ket{\vec p\,} =\vec\nabla \left( e^{i\phi(\vec x)}  \ket{\vec p\,}_a \right)
        \nn\\
	&= \left( i \vec\nabla 
    \phi(\vec x\,)  \right)   
    e^{i\phi(\vec x)}  
 \ket{\vec p\,}_a + i  
    e^{i\phi(\vec x)}  
    \vec u \ket{\vec p\,}_a  .
\end{align}
Hence need
\be
 \vec\nabla \phi(\vec x\,) 
    + \vec u\,(\vec x\,) = 0 .
\ee
The solution is
\be
\phi(\vec x\,) = -  \int^x d\vec x\,' 
    \cdot \vec u\,(\vec x\,') 		.
\ee
Since $\phi(\vec x\,)$ can be found, the state with $(\eth/\eth x_i) \ket{p} = 0$ exists. \hfill 

This completes the demonstration of the Lemma, and so also completes the demonstration that a canonical center of mass coordinate can be defined if one has momentum eigenstates.

\bibliography{hadchdens}

\begin{thebibliography}{13}%
\makeatletter
\providecommand \@ifxundefined [1]{%
 \@ifx{#1\undefined}
}%
\providecommand \@ifnum [1]{%
 \ifnum #1\expandafter \@firstoftwo
 \else \expandafter \@secondoftwo
 \fi
}%
\providecommand \@ifx [1]{%
 \ifx #1\expandafter \@firstoftwo
 \else \expandafter \@secondoftwo
 \fi
}%
\providecommand \natexlab [1]{#1}%
\providecommand \enquote  [1]{``#1''}%
\providecommand \bibnamefont  [1]{#1}%
\providecommand \bibfnamefont [1]{#1}%
\providecommand \citenamefont [1]{#1}%
\providecommand \href@noop [0]{\@secondoftwo}%
\providecommand \href [0]{\begingroup \@sanitize@url \@href}%
\providecommand \@href[1]{\@@startlink{#1}\@@href}%
\providecommand \@@href[1]{\endgroup#1\@@endlink}%
\providecommand \@sanitize@url [0]{\catcode `\\12\catcode `\$12\catcode
  `\&12\catcode `\#12\catcode `\^12\catcode `\_12\catcode `\%12\relax}%
\providecommand \@@startlink[1]{}%
\providecommand \@@endlink[0]{}%
\providecommand \url  [0]{\begingroup\@sanitize@url \@url }%
\providecommand \@url [1]{\endgroup\@href {#1}{\urlprefix }}%
\providecommand \urlprefix  [0]{URL }%
\providecommand \Eprint [0]{\href }%
\providecommand \doibase [0]{http://dx.doi.org/}%
\providecommand \selectlanguage [0]{\@gobble}%
\providecommand \bibinfo  [0]{\@secondoftwo}%
\providecommand \bibfield  [0]{\@secondoftwo}%
\providecommand \translation [1]{[#1]}%
\providecommand \BibitemOpen [0]{}%
\providecommand \bibitemStop [0]{}%
\providecommand \bibitemNoStop [0]{.\EOS\space}%
\providecommand \EOS [0]{\spacefactor3000\relax}%
\providecommand \BibitemShut  [1]{\csname bibitem#1\endcsname}%
\let\auto@bib@innerbib\@empty
\bibitem [{\citenamefont {Miller}(2007)}]{Miller:2007uy}%
  \BibitemOpen
  \bibfield  {author} {\bibinfo {author} {\bibfnamefont {G.~A.}\ \bibnamefont
  {Miller}},\ }\href {\doibase 10.1103/PhysRevLett.99.112001} {\bibfield
  {journal} {\bibinfo  {journal} {Phys. Rev. Lett.}\ }\textbf {\bibinfo
  {volume} {99}},\ \bibinfo {pages} {112001} (\bibinfo {year} {2007})},\
  \Eprint {http://arxiv.org/abs/0705.2409} {arXiv:0705.2409 [nucl-th]}
  \BibitemShut {NoStop}%
\bibitem [{\citenamefont {Burkardt}(2003)}]{Burkardt:2002hr}%
  \BibitemOpen
  \bibfield  {author} {\bibinfo {author} {\bibfnamefont {M.}~\bibnamefont
  {Burkardt}},\ }\href {\doibase 10.1142/S0217751X03012370} {\bibfield
  {journal} {\bibinfo  {journal} {Int. J. Mod. Phys. A}\ }\textbf {\bibinfo
  {volume} {18}},\ \bibinfo {pages} {173} (\bibinfo {year} {2003})},\ \Eprint
  {http://arxiv.org/abs/hep-ph/0207047} {arXiv:hep-ph/0207047} \BibitemShut
  {NoStop}%
\bibitem [{\citenamefont {Epelbaum}\ \emph {et~al.}(2022)\citenamefont
  {Epelbaum}, \citenamefont {Gegelia}, \citenamefont {Lange}, \citenamefont
  {Mei\ss{}ner},\ and\ \citenamefont {Polyakov}}]{Epelbaum:2022fjc}%
  \BibitemOpen
  \bibfield  {author} {\bibinfo {author} {\bibfnamefont {E.}~\bibnamefont
  {Epelbaum}}, \bibinfo {author} {\bibfnamefont {J.}~\bibnamefont {Gegelia}},
  \bibinfo {author} {\bibfnamefont {N.}~\bibnamefont {Lange}}, \bibinfo
  {author} {\bibfnamefont {U.~G.}\ \bibnamefont {Mei\ss{}ner}}, \ and\ \bibinfo
  {author} {\bibfnamefont {M.~V.}\ \bibnamefont {Polyakov}},\ }\href {\doibase
  10.1103/PhysRevLett.129.012001} {\bibfield  {journal} {\bibinfo  {journal}
  {Phys. Rev. Lett.}\ }\textbf {\bibinfo {volume} {129}},\ \bibinfo {pages}
  {012001} (\bibinfo {year} {2022})},\ \Eprint
  {http://arxiv.org/abs/2201.02565} {arXiv:2201.02565 [hep-ph]} \BibitemShut
  {NoStop}%
\bibitem [{\citenamefont {Panteleeva}\ \emph {et~al.}(2022)\citenamefont
  {Panteleeva}, \citenamefont {Epelbaum}, \citenamefont {Gegelia},\ and\
  \citenamefont {Mei\ss{}ner}}]{Panteleeva:2022khw}%
  \BibitemOpen
  \bibfield  {author} {\bibinfo {author} {\bibfnamefont {J.~Y.}\ \bibnamefont
  {Panteleeva}}, \bibinfo {author} {\bibfnamefont {E.}~\bibnamefont
  {Epelbaum}}, \bibinfo {author} {\bibfnamefont {J.}~\bibnamefont {Gegelia}}, \
  and\ \bibinfo {author} {\bibfnamefont {U.~G.}\ \bibnamefont {Mei\ss{}ner}},\
  }\href {\doibase 10.1103/PhysRevD.106.056019} {\bibfield  {journal} {\bibinfo
   {journal} {Phys. Rev. D}\ }\textbf {\bibinfo {volume} {106}},\ \bibinfo
  {pages} {056019} (\bibinfo {year} {2022})},\ \Eprint
  {http://arxiv.org/abs/2205.15061} {arXiv:2205.15061 [hep-ph]} \BibitemShut
  {NoStop}%
\bibitem [{\citenamefont {Panteleeva}\ \emph
  {et~al.}(2023{\natexlab{a}})\citenamefont {Panteleeva}, \citenamefont
  {Epelbaum}, \citenamefont {Gegelia},\ and\ \citenamefont
  {Mei\ss{}ner}}]{Panteleeva:2022uii}%
  \BibitemOpen
  \bibfield  {author} {\bibinfo {author} {\bibfnamefont {J.~Y.}\ \bibnamefont
  {Panteleeva}}, \bibinfo {author} {\bibfnamefont {E.}~\bibnamefont
  {Epelbaum}}, \bibinfo {author} {\bibfnamefont {J.}~\bibnamefont {Gegelia}}, \
  and\ \bibinfo {author} {\bibfnamefont {U.~G.}\ \bibnamefont {Mei\ss{}ner}},\
  }\href {\doibase 10.1140/epjc/s10052-023-11746-x} {\bibfield  {journal}
  {\bibinfo  {journal} {Eur. Phys. J. C}\ }\textbf {\bibinfo {volume} {83}},\
  \bibinfo {pages} {617} (\bibinfo {year} {2023}{\natexlab{a}})},\ \Eprint
  {http://arxiv.org/abs/2211.09596} {arXiv:2211.09596 [hep-ph]} \BibitemShut
  {NoStop}%
\bibitem [{\citenamefont {Panteleeva}\ \emph
  {et~al.}(2023{\natexlab{b}})\citenamefont {Panteleeva}, \citenamefont
  {Epelbaum}, \citenamefont {Gegelia},\ and\ \citenamefont
  {Mei\ss{}ner}}]{Panteleeva:2023evj}%
  \BibitemOpen
  \bibfield  {author} {\bibinfo {author} {\bibfnamefont {J.~Y.}\ \bibnamefont
  {Panteleeva}}, \bibinfo {author} {\bibfnamefont {E.}~\bibnamefont
  {Epelbaum}}, \bibinfo {author} {\bibfnamefont {J.}~\bibnamefont {Gegelia}}, \
  and\ \bibinfo {author} {\bibfnamefont {U.~G.}\ \bibnamefont {Mei\ss{}ner}},\
  }\href {\doibase 10.1007/JHEP07(2023)237} {\bibfield  {journal} {\bibinfo
  {journal} {JHEP}\ }\textbf {\bibinfo {volume} {07}},\ \bibinfo {pages} {237}
  (\bibinfo {year} {2023}{\natexlab{b}})},\ \Eprint
  {http://arxiv.org/abs/2305.01491} {arXiv:2305.01491 [hep-ph]} \BibitemShut
  {NoStop}%
\bibitem [{\citenamefont {Freese}\ and\ \citenamefont
  {Miller}(2022)}]{Freese:2021mzg}%
  \BibitemOpen
  \bibfield  {author} {\bibinfo {author} {\bibfnamefont {A.}~\bibnamefont
  {Freese}}\ and\ \bibinfo {author} {\bibfnamefont {G.~A.}\ \bibnamefont
  {Miller}},\ }\href {\doibase 10.1103/PhysRevD.105.014003} {\bibfield
  {journal} {\bibinfo  {journal} {Phys. Rev. D}\ }\textbf {\bibinfo {volume}
  {105}},\ \bibinfo {pages} {014003} (\bibinfo {year} {2022})},\ \Eprint
  {http://arxiv.org/abs/2108.03301} {arXiv:2108.03301 [hep-ph]} \BibitemShut
  {NoStop}%
\bibitem [{\citenamefont {Jaffe}(2021)}]{Jaffe:2020ebz}%
  \BibitemOpen
  \bibfield  {author} {\bibinfo {author} {\bibfnamefont {R.~L.}\ \bibnamefont
  {Jaffe}},\ }\href {\doibase 10.1103/PhysRevD.103.016017} {\bibfield
  {journal} {\bibinfo  {journal} {Phys. Rev. D}\ }\textbf {\bibinfo {volume}
  {103}},\ \bibinfo {pages} {016017} (\bibinfo {year} {2021})},\ \Eprint
  {http://arxiv.org/abs/2010.15887} {arXiv:2010.15887 [hep-ph]} \BibitemShut
  {NoStop}%
\bibitem [{\citenamefont {Lorc\'e}\ \emph {et~al.}(2022)\citenamefont
  {Lorc\'e}, \citenamefont {Schweitzer},\ and\ \citenamefont
  {Tezgin}}]{Lorce:2022cle}%
  \BibitemOpen
  \bibfield  {author} {\bibinfo {author} {\bibfnamefont {C.}~\bibnamefont
  {Lorc\'e}}, \bibinfo {author} {\bibfnamefont {P.}~\bibnamefont {Schweitzer}},
  \ and\ \bibinfo {author} {\bibfnamefont {K.}~\bibnamefont {Tezgin}},\ }\href
  {\doibase 10.1103/PhysRevD.106.014012} {\bibfield  {journal} {\bibinfo
  {journal} {Phys. Rev. D}\ }\textbf {\bibinfo {volume} {106}},\ \bibinfo
  {pages} {014012} (\bibinfo {year} {2022})},\ \Eprint
  {http://arxiv.org/abs/2202.01192} {arXiv:2202.01192 [hep-ph]} \BibitemShut
  {NoStop}%
\bibitem [{\citenamefont {Guo}\ \emph {et~al.}(2021)\citenamefont {Guo},
  \citenamefont {Ji},\ and\ \citenamefont {Shiells}}]{Guo:2021aik}%
  \BibitemOpen
  \bibfield  {author} {\bibinfo {author} {\bibfnamefont {Y.}~\bibnamefont
  {Guo}}, \bibinfo {author} {\bibfnamefont {X.}~\bibnamefont {Ji}}, \ and\
  \bibinfo {author} {\bibfnamefont {K.}~\bibnamefont {Shiells}},\ }\href
  {\doibase 10.1016/j.nuclphysb.2021.115440} {\bibfield  {journal} {\bibinfo
  {journal} {Nucl. Phys. B}\ }\textbf {\bibinfo {volume} {969}},\ \bibinfo
  {pages} {115440} (\bibinfo {year} {2021})},\ \Eprint
  {http://arxiv.org/abs/2101.05243} {arXiv:2101.05243 [hep-ph]} \BibitemShut
  {NoStop}%
\bibitem [{\citenamefont {Carlson}\ and\ \citenamefont
  {Vanderhaeghen}(2008)}]{Carlson:2007xd}%
  \BibitemOpen
  \bibfield  {author} {\bibinfo {author} {\bibfnamefont {C.~E.}\ \bibnamefont
  {Carlson}}\ and\ \bibinfo {author} {\bibfnamefont {M.}~\bibnamefont
  {Vanderhaeghen}},\ }\href {\doibase 10.1103/PhysRevLett.100.032004}
  {\bibfield  {journal} {\bibinfo  {journal} {Phys. Rev. Lett.}\ }\textbf
  {\bibinfo {volume} {100}},\ \bibinfo {pages} {032004} (\bibinfo {year}
  {2008})},\ \Eprint {http://arxiv.org/abs/0710.0835} {arXiv:0710.0835
  [hep-ph]} \BibitemShut {NoStop}%
\bibitem [{\citenamefont {Miller}(2010)}]{Miller:2010nz}%
  \BibitemOpen
  \bibfield  {author} {\bibinfo {author} {\bibfnamefont {G.~A.}\ \bibnamefont
  {Miller}},\ }\href {\doibase 10.1146/annurev.nucl.012809.104508} {\bibfield
  {journal} {\bibinfo  {journal} {Ann. Rev. Nucl. Part. Sci.}\ }\textbf
  {\bibinfo {volume} {60}},\ \bibinfo {pages} {1} (\bibinfo {year} {2010})},\
  \Eprint {http://arxiv.org/abs/1002.0355} {arXiv:1002.0355 [nucl-th]}
  \BibitemShut {NoStop}%
\bibitem [{\citenamefont {Pryce}(1948)}]{Pryce:1948pf}%
  \BibitemOpen
  \bibfield  {author} {\bibinfo {author} {\bibfnamefont {M.~H.~L.}\
  \bibnamefont {Pryce}},\ }\href {\doibase 10.1098/rspa.1948.0103} {\bibfield
  {journal} {\bibinfo  {journal} {Proc. Roy. Soc. Lond. A}\ }\textbf {\bibinfo
  {volume} {195}},\ \bibinfo {pages} {62} (\bibinfo {year} {1948})}\BibitemShut
  {NoStop}%
\end{thebibliography}%

\end{document}